\documentclass[twocolumn,showpacs,showkeys,preprintnumbers,amsmath,amssymb]{revtex4-1}
\usepackage{bbm}
\usepackage{amsfonts}

\usepackage{epsfig}
\usepackage{graphicx}
\usepackage{amssymb}   
\usepackage{dcolumn}
\usepackage{bm}
\usepackage[colorlinks,citecolor=blue, linkcolor=blue,hyperindex,CJKbookmarks]{hyperref}
\begin{document}

\title{Proposal for Observing Dynamic Jahn-Teller Effect of Single Solid-State Defects}

\author{Xing Xiao$^{1,2}$}
\author{Nan Zhao$^{1,3}$}
\altaffiliation{nzhao@csrc.ac.cn}
\affiliation{$^{1}$Beijing Computational Science Research Center, Beijing 100084, China\\
$^{2}$College of Physics and Electronic Information,
Gannan Normal University, Ganzhou 341000, China\\
$^{3}$Synergetic Innovation Center of Quantum Information and Quantum
Physics, University of Science and Technology of China, Hefei, Anhui
230026, China}

\begin{abstract}
Jahn-Teller effect (JTE) widely exists in polyatomic systems including organic molecules, nano-magnets, and solid-state defects.
Detecting the JTE at single-molecule level can provide unique properties about the detected individual object.
However,  such measurements are challenging because of the weak signals associated with a single quantum object.
Here, we propose that the dynamic JTE of single defects in solids can be observed with nearby quantum sensors.
With numerical simulations, we demonstrate the real-time monitoring of quantum jumps between different stable configurations of single substitutional nitrogen defect centers (P1 centers) in diamond. This is achieved by measuring the spin coherence of a single nitrogen-vacancy (NV) center near the P1 center with the double electron-electron resonance (DEER) technique.
Our work extends the ability of NV center as a quantum probe to sense the rich physics in various electron-vibrational coupled systems.
\end{abstract}
\maketitle

\section{Introduction-.}
The Jahn-Teller effect (JTE) is one of the most important phenomena caused by electron-vibrational interaction,
which has broad impact on both physics and chemistry.
In fundamental physics, JTE relates to the symmetry-breaking concept, where the system Hamiltonian has a certain symmetry, but the ground state does not.
In chemistry and condensed matter physics, the JTE is essential in understanding the structure, the optical and magnetic properties of polyatomic systems like molecules and solid state point defects.

The dynamic JTE describes the transitions of the system from one stable configuration to another.
This configuration transition can happen with the help of thermal excitation or quantum tunneling \cite{Bersuker2006}.
The dynamic JTE is usually measured by the change of optical or magnetic resonance properties under different conditions (e.g., temperature and strain).
The characteristic parameters of the JTE, including the the potential barrier height and the tunneling rate, are usually inferred indirectly from the ensemble averaged quantities such as optical and magnetic transition frequencies and line widths.
Directly monitoring the quantum jumps of individual systems in real-time (e.g., a single molecule or a single solid-state defect) is intriguing and will provide more knowledges about the dynamic JTE and the local environment.
However, to our knowledge, the real-time measurement of dynamic JTE of an individual system is not achieved because of the weak signal associated with a single molecule or a single defect.

In this work, we propose to measure the dynamic JTE of a typical kind of single solid-state defects, namely, the substitutional nitrogen defect centers (P1 centers)  in diamond \cite{Davies1981}.
Although the undistorted structure has a tetrahedral symmetry, the energetically stable configurations of P1 center have triangle symmetry due to the Jahn-Teller distortion.
There are four equivalent orientations of P1 centers corresponding to the nitrogen atom shifting along the direction of four N-C bonds.
The JTE of P1 centers was observed via electron spin resonance \cite{Smith1959} and electron-nuclear double resonance \cite{Cook1966,Cox1994}, and the orientation relaxation (reorientation) rate was measured and calculated in a wide range of temperature.
At temperature $T>250~{\rm K}$, the reorientation rate $\nu$ follows the Arrhenius law \cite{Davies1981,Ammerlaan1981}
\begin{equation}
\label{EQ:1}
\nu=\nu_0 \exp\left[-\frac{V}{k_{\rm B}T}\right],
\end{equation}
where $k_{\rm B}$ is the Boltzmann constant, $\nu_0\sim 4\times10^{12}~{\rm s}^{-1}$ and $V=0.76~{\rm eV}$.
At low temperature ( $T\lesssim 200~{\rm K}$), the reorientation rate deviates from the Arrhenius law, ranging from  $10^{-3}~{\rm s}^{-1}$ to $10^{-5}~{\rm s}^{-1}$ \cite{Ammerlaan1981}.
The low rate at low temperature allows us to observe the reorientation process of individual P1 centers in real-time.

We propose to monitor the reorientation process of single P1 centers using the nitrogen-vacancy (NV) centers in diamond as quantum probe.
The NV centers have been demonstrated to be ultra-sensitive magnetometers with atomic-scale resolution \cite{Balasubramanian2008,Maze2008}.
Particularly, single NV centers were used to detect the weak magnetic signals emitted form single nuclear spin clusters \cite{Kolkowitz2012a,Taminiau2012a,Zhao2011a,Zhao2012,Mamin2013,Staudacher2013,Muller2014}.
Also, non-magnetic signals can be detected by converting to magnetic ones \cite{Cai2014}.
Notice that the Jahn-Teller distortion modifies the magnetic resonance frequency of the P1 centers electron spins \cite{Hanson2008,Lange2010}, and the spins couples to the NV center electron spin through the magnetic dipolar interaction.
Thus, it is possible to readout the P1 center state via an adjacent NV center.

The readout of the an individual P1 center orientation can be realized by using the double electron-electron resonance (DEER) technique \cite{Neumann2010a,Shi2013a}.
It is well-established that high concentration P1 centers in diamond (e.g., $\sim 10^2$~ppm) serve as an electron spin bath, which causes the electron spin decoherence of the NV centers \cite{Hanson2008,Lange2010}.
The decoherence effect can be partially removed by resonantly driving the P1 center bath spins using microwave pulses \cite{DeLange2012}.
Due to the hyperfine interaction to the nitrogen nuclear spins, the P1 centers with different orientations can have different magnetic resonance frequencies in strong magnetic fields \cite{Cook1966, Hanson2008,DeLange2012}.
This enables driving P1 centers with particular orientation using frequency-selective microwave pulses \cite{DeLange2012}.
In the following we show that, among a large number of P1 center bath spins, the nearest P1 center to the NV center usually has much more significant impact on the NV center spin coherence, whose orientation can be readout in a single-shot manner \cite{Neumann2010} by repetitive measurement on the NV center.

The proposed method is not limited in specific detected systems and, in principle, can be generalized to measure the JTE of other single quantum systems.
Possible application includes monitoring the dynamic JTE of single molecular nano-magnet with shallow NV centers in diamond.
This work extends the physical processes that NV centers can detect, and makes the investigation of JTE at single-molecule level possible.

\section{Observing dynamics Jahn-Teller effect}
\subsection{Double Electron-Electron Resonance}

\begin{figure}[t]
  \includegraphics[width=0.5\textwidth]{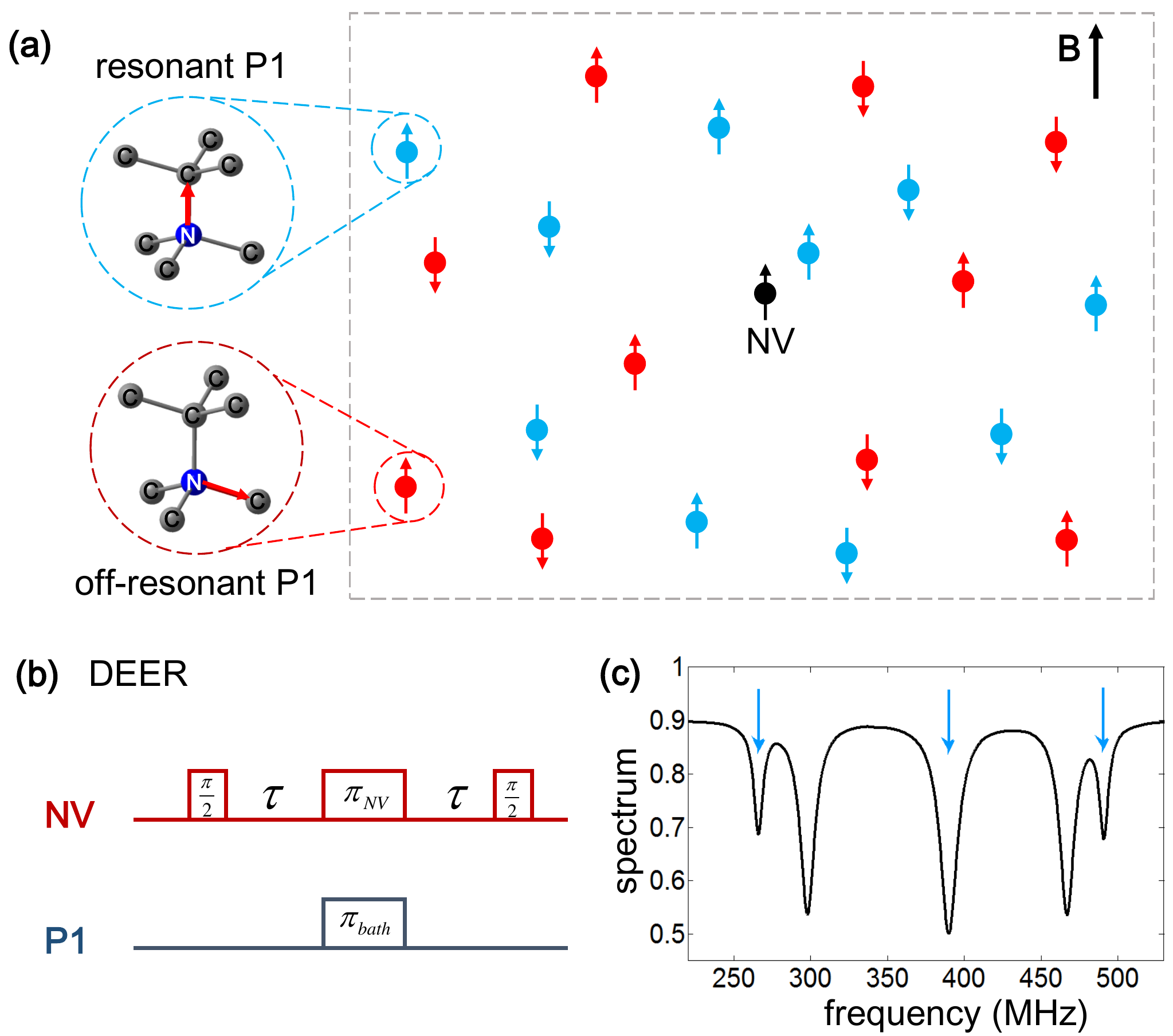}
\caption{(color online) (a) Schematic illustration of NV center (the black spin) in an electron spin bath consisting of P1 centers (the red and blue spins) .
The structure of P1 centers is shown on the left, and the red arrow indicates the elongated N-C bond.
 In a strong magnetic field $\mathbf{B}$ along the $[111]$ crystallographic direction (the $z$-axis),
 the bath spins are randomly aligned parallel or anti-parallel in the $z$ direction.
 All the bath spins are classified as the resonant group (blue) and the off-resonant group (red) according to their magnetic transition frequencies (see text).
(b) The double electron-electron resonance (DEER) pulse sequence.
(c) Magnetic resonance spectroscopy of P1 centers. Five dips associate with different orientation $v$ and $^{14}$N nuclear spin state $I_z$.
In this paper, we focus on driving the P1 centers with the 3 resonant frequencies indicated by the arrows.
}
\label{FIG:1}
\end{figure}

We consider a type-Ib diamond sample, where single NV centers are embedded in the electron spin bath of P1 centers (see Fig.~\ref{FIG:1}).
In a strong magnetic field $\mathbf{B}$ (e.g., $B>200~{\rm Gauss}$) along the NV axis (assumed to be the $z$-axis), the NV center spin $\mathbf{S}_0$ couples to $N$ P1 center bath spins $\mathbf{S}_k$ (for $k=1\dots N$) via dipolar interaction, and the Hamiltonian reads
\begin{equation}
\label{EQ:Hamiltonian}
H=D \left(S_0^z\right)^2 -\gamma B \sum_{k=0}^N S_k^{z} + S_0^z \sum_{k=1}^N b_k S_k^{z},
\end{equation}
where $D=2.87~{\rm GHz}$ is the zero field splitting of the NV center,  $\gamma$ is the gyromagnetic ratio of electron spins, and the last term describes the dipolar interaction in strong field with $b_k= \frac{\mu_0\gamma^2}{4\pi r_k^3}\left[1-3 (n_k^z)^2\right]$ ($\mu_0$ is the vacuum permeability, and $n_k^z$ is the directional cosine of the $k$th bath spin) \cite{Hanson2008}.
In Eq.~(\ref{EQ:Hamiltonian}), we have omitted the dipolar coupling between bath spins, since it has negligible effect in the short time scale ($\sim {\rm \mu s}$) we are interested in.

For a typical concentration $c=200~{\rm ppm}$ of P1 centers, the NV center electron spin coherence decays in about $T_2^{*}\lesssim 1~{\rm \mu s}$ (i.e., the free-induction decay, or, FID) due to the noise field created by the bath spins (i.e. the inhomogeneous broadening).
With the well-known spin-echo technique, where a $\pi$-pulse is applied on the NV center at time $t=\tau$,
the static fluctuations will be refocused and the coherence recovers at time $t=2\tau$.
In this case, the coherence time is extended to $T_2$ which is much longer than $T_2^{*}$ \cite{Hanson2008}.

The DEER sequence uses an additional $\pi$-pulse to flip the P1 center spins at $t=\tau$.
The resonant frequency of the P1 center depends on its orientations and nuclear spin states \cite{Cook1966, Hanson2008,DeLange2012}.
In strong magnetic fields, the electron spin of the P1 center couples to its nitrogen nuclear spin by the hyperfine interaction
\begin{equation}
H_{\rm hf}= \sum_{k=1}^{N} A_k^{(v)} S_k^z I_k^z,
\end{equation}
where the coupling strength $A_k^{(v)}$ depends on the P1 center orientation $v\in\{a, b, c, d\}$.
For the P1 centers with distortion axis parallel with the magnetic field direction (the $v=a$ case), the coupling strength $A_k^{(a)}=114~{\rm MHz}$.
Otherwise, if the distortion axis lies in the other three equivalent directions (i.e.,  $v=b, c$ or $d$), the coupling strength $A_k^{(v)}=86~{\rm MHz}$.
For a given orientation $v$, the hyperfine coupling results in  three resonant peaks corresponding to nuclear spin magnetic quantum number $I_z=0$ and $\pm 1$.
The resonant frequencies for $I_z=0$ are degenerate for all four orientations, while the parallel orientation (the $v=a$ case) has larger splitting for the  $I_z=\pm 1$ resonant frequencies (see Fig.~\ref{FIG:1}c).

The effect of the two $\pi$-pulses cancels each other if both NV center and P1 centers are resonantly flipped.
The resultant pulse sequence, in this case,  is equivalent to the FID case (with two $\pi/2$-pulses only).
Thus, the microwave $\pi$-pulse on P1 centers divides the bath spins into two groups (see Fig.~\ref{FIG:1}): (i) the resonant group $G_{\rm res}$, in which the spins are flipped by the pulse; and (ii) the off-resonant group $G_{\rm off}$, in which the spins are unaffected.
The P1 centers in the off-resonant group contribute little to the NV center spin decoherence due to the refocusing pulse on NV center, while the bath spins in the resonant group, as long as the spin number $\vert G_{\rm res}\vert$ is not too small, dominate the decoherence. In this case,  the coherence decays as (see Appendix \ref{appendix:B})
\begin{equation}
L_{\rm DEER}(t)\approx\prod_{k \in G_{\rm res}} \cos \left(\frac{b_k t}{2}\right),
\end{equation}
where the product is performed over all the P1 centers in the resonant group.

\begin{figure}
  \includegraphics[width=0.5\textwidth]{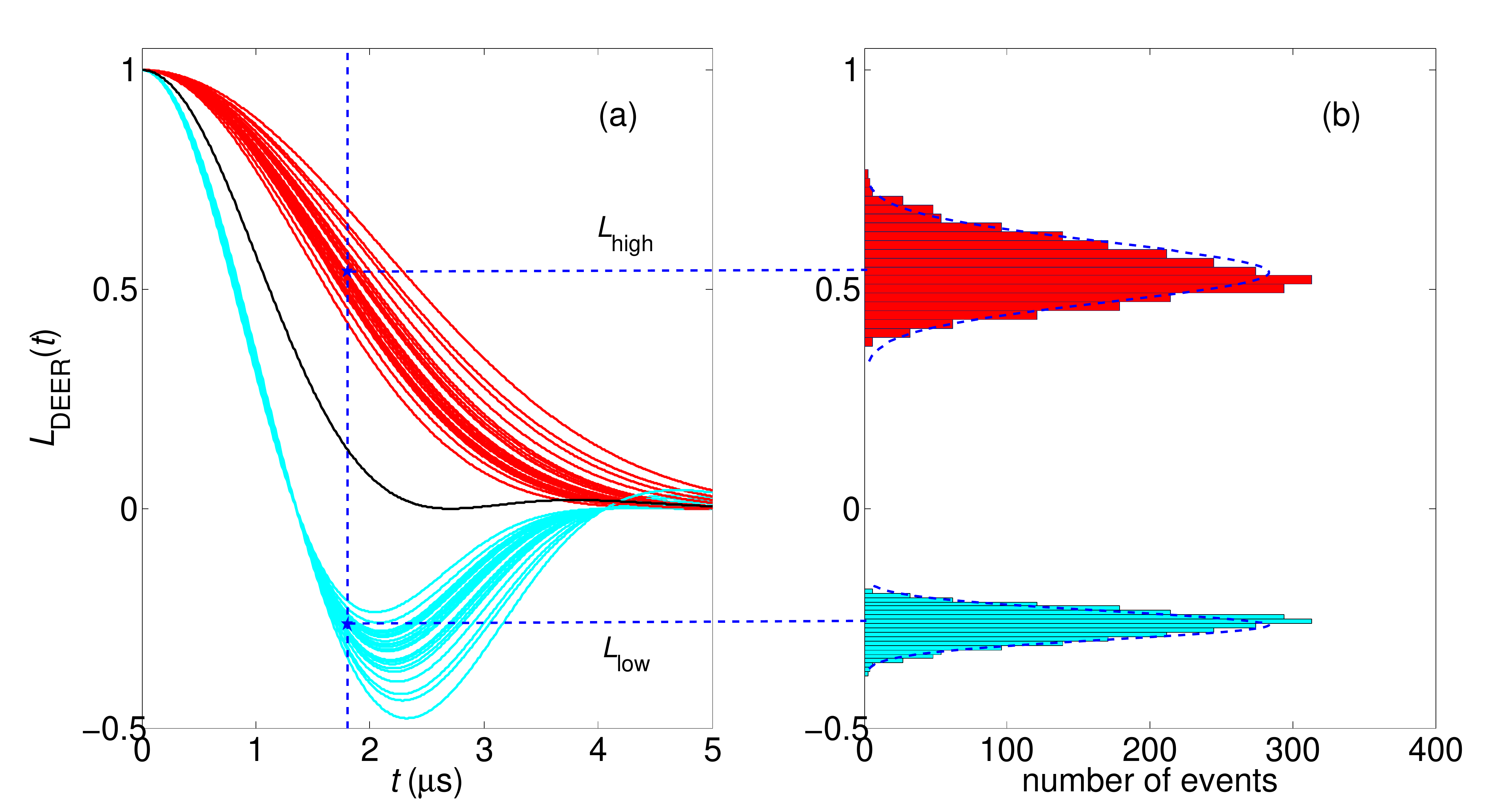}
\caption{(color online) (a) NV center electron spin coherence under DEER sequence. Each curve is calculated with randomly generated P1 center orientations. The orientation is assumed to be unchanged (static) during the coherent evolution (e.g., $t<5~{\rm \mu s}$).
The curves in cyan (red) are the NV center coherence with the nearest P1 center (i.e., the most strongly coupled one) in the resonant (off-resonant) group.
The black solid curve is the NV center coherence averaged over all the P1 center orientations.
The vertical dashed line indicates the working point where the single-shot readout is performed.
The parameter $\eta_1$ of this bath configuration is $\eta_1=0.64$.
(b) Histogram of the NV center spin coherence at $t=1.8~{\rm \mu s}$ (the working point) for different bath spin orientations.}
\label{FIG:2}
\end{figure}

By setting excitation frequencies of the microwave pulses, we can choose the orientation and nuclear spin states, in which the P1 centers contribute to the NV center decoherence.
To be concrete, we consider the microwave pulses drive the three peaks separated by $114~{\rm MHz}$ (see Fig.~\ref{FIG:1}c).
In this case, the resonant group $G_{\rm res}$ contains the P1 centers in 6 states $\{v, I_z\}=\{a, \pm 1\}$ and $\{v, 0\}$, for $v=a,b,c$ and $d$.
Assuming that the 12 states of P1 centers $\{v, I_z\}$ are randomly populated with equal probability, we have about half P1 centers belonging to the resonant group $G_{\rm res}$.

\subsection{Single-Shot Readout of P1 Center Orientation}
Because of the inverse-cubic dependence of the dipolar coupling strength on distance, the adjacent P1 centers to the NV centers have much more significant contributions to the spin coherence.
For the moment, we focus on the nearest P1 center to the NV.
In the case of P1 center concentration $c=200~{\rm ppm}$, the typical distance between the NV center and the nearest P1 center is several nanometers, and the coupling strength, denoted by $b_1$,  is in the order of $\sim {\rm MHz}$.
To quantify the contribution of the nearest P1 center, we define the ratio
\begin{equation}
\eta_1 = \frac{b_1^2 }{\sum_{k=1}^N b_k^{2}},
\label{ratio}
\end{equation}
where $b_k^2$ characterize the size of the fluctuation due to the $k$th P1 center.
Our simulation shows that the probability of strongly coupled P1 center is not too small.
About $13\%$ randomly generated bath configurations have the ratio $\eta_1 >  0.5$ (see Appendix \ref{appendix:A}).

For a given configuration with large $\eta_1$ (see Fig.~\ref{FIG:2}), the NV center decoherence behavior in time domain strongly depends on whether or not the the nearest P1 center is in the resonant group $G_{\rm res}$.
Figure~\ref{FIG:2} shows the calculated NV center coherence under DEER sequence for a given bath configuration and with random states $\{v, I_z\}$ being assigned to each P1 centers.
The decoherence behavior dramatically changes if the state of the nearest P1 center changes from the resonant group to the off-resonant group, while the state change of other P1 centers only causes small modifications.
By choosing appropriate working point (the vertical dashed line shown in Fig.~\ref{FIG:2}a), one can realize the single-shot readout of the P1 center state by repetitive measurement on the NV centers.

\begin{figure*}
  \includegraphics[width=1\textwidth]{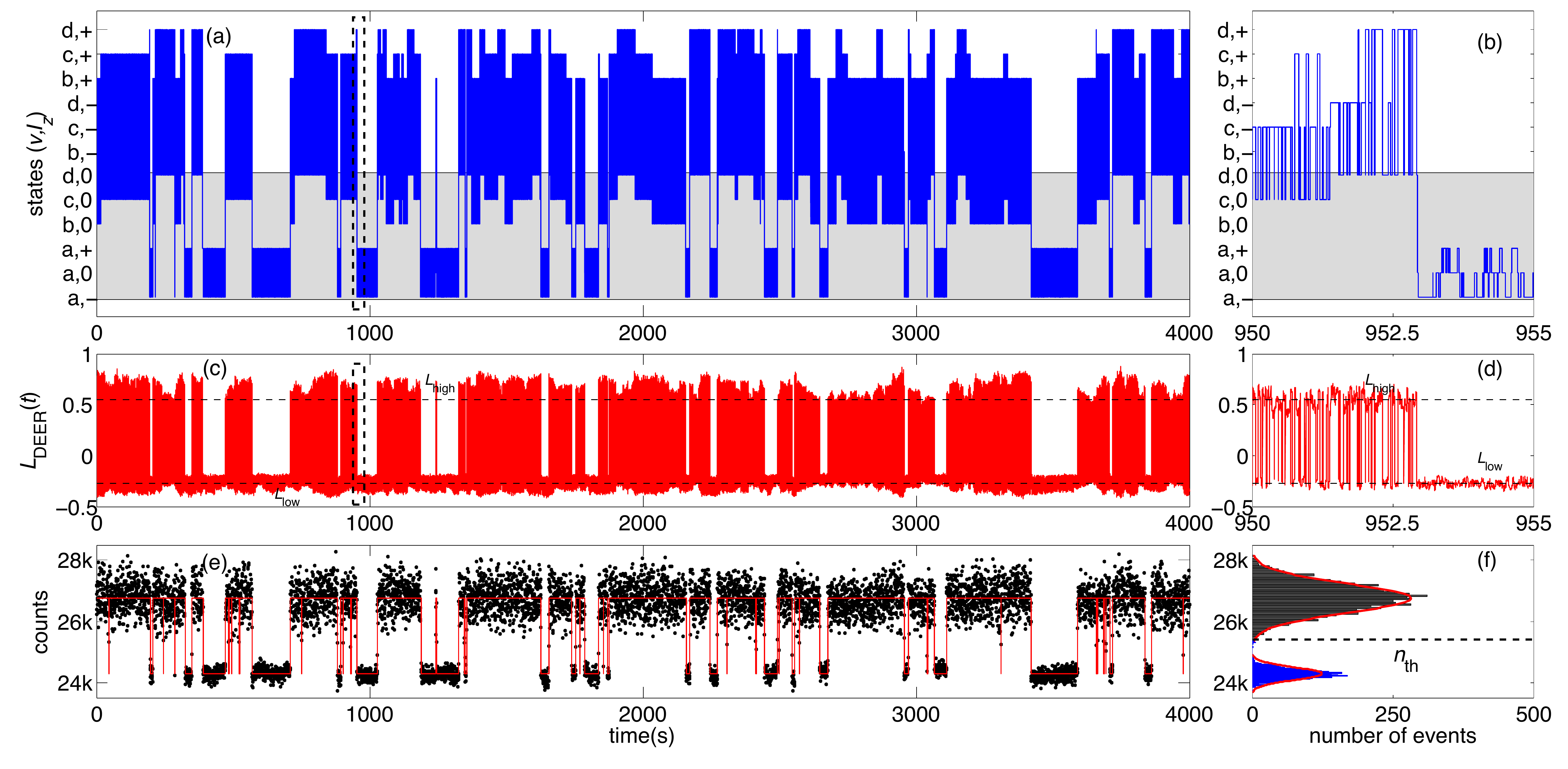}
\caption{(color online) (a) State jumps of the nearest P1 center to the NV center. The simulation is performed with orientation relaxation time $\tau_{v}=100~{\rm s}$ (at $T=262~{\rm K}$) and nuclear spin relaxation time $T_{1 n}=50~{\rm ms}$.
(b) Zoom-in of the boxed region in (a).
(c) The calculated NV center spin coherence (at  the working point $t=1.8~{\rm \mu s}$) according to the P1 center states (see Fig.~\ref{FIG:2}a).
(d) Zoom-in of the boxed region in (c).
(e) Simulated quantum jumps of the P1 center orientations. Each data point corresponds to the number of photons collected in $0.5~{\rm s}$ ($10^5$ repetitions of the DEER sequence).
(f) Histogram of the photons collected in $0.5~{\rm s}$. The red curve is the double-Gaussian fitting. The dashed line indicates the threshold value $n_{\rm th}$.}
\label{FIG:3}       
\end{figure*}

\subsection{Monitoring Quantum Jump in Real-Time}
At finite temperature, both the P1 center orientation and its nuclear spin state are changing in time.
The change of P1 center state is simulated by a Markov stochastic process (see the Appendix \ref{appendix:C}).
Figure~\ref{FIG:3}a shows a typical realization of the jump process of the nearest P1 center.
The P1 centers located at different positions have similar random jump behavior (not shown).
In Fig.~\ref{FIG:3}, we use the orientation relaxation time $\tau_v=100~{\rm s}$ corresponding to the temperature $T=262~{\rm K}$ [see Eq.~(\ref{EQ:1})].
The nitrogen nuclear spin life time $T_{1n}$ is much shorter, assumed to be  $50~{\rm ms}$ \cite{Note}.

The quantum jump of P1 center can be monitored by measuring the NV center coherence with DEER sequence.
A single DEER sequence is completed in about $T_{\rm DEER}=5~{\rm \mu s}$,
which consists of the time for laser initialization/readout of NV center spin state ($3~{\rm \mu s}$), microwave pulse duration ($\sim 10^2~{\rm ns}$), and the time for coherent evolution ($1-2~{\rm \mu s}$ according to the working point, see Fig.~\ref{FIG:2}a).
During the time of a single DEER sequence $T_{\rm DEER}$, the P1 center state is hardly changed (i.e., $T_{\rm DEER}\ll T_{1n}$ and $\tau_v$).
Figure~\ref{FIG:3}c shows the evolution of the NV center spin coherence, which is calculated according to the P1 center states at each instant.
The spin coherence switches between two values $L_{\rm low}$ and $L_{\rm high}$ whenever the nearest P1 center jumps into or out of the resonant group $G_{\rm res}$.
The state change of those P1 centers with much weaker coupling causes the small fluctuation of NV center coherence around  $L_{\rm low}$ or $L_{\rm high}$.

In realistic measurements, one has to repeat the DEER sequence, e.g., $M=10^5$ times, to build up statistics.
The total measurement time $T_M=MT_{\rm DEER}=0.5~{\rm s}$ is much longer than the nuclear spin relaxation time $T_{1n}$,
but can be much shorter than the orientation relaxation time $\tau_v$ at low temperature.
By recording the number of photons collected in every $0.5~{\rm s}$, one averages out the coherence change due to nuclear spin flipping events, leaving only the reorientation events being monitored.

Figure~\ref{FIG:3}e shows a numerical simulation of the quantum jump process. Each data point represents the photon number collected in $0.5~{\rm s}$.
Higher counts correspond to the nearest P1 center in the $v=b, c, $ or $d$ orientations, while the lower counts indicate that its orientation is along the $z$ axis ($v=a$).
With the strongly coupled nearest P1 center and an appropriately chosen working point, the photon counts of every $0.5~{\rm s}$ follow a well-separated double-Gaussian distribution (see Fig.~\ref{FIG:3}f).
The peak separation (the contrast) is determined by the coherence difference $L_{\rm high}-L_{\rm low}$ at the working point.
The peak widths (photon number fluctuations) come from the photon shot noise, the fluctuation of spin coherence around $L_{\rm low}$ and $L_{\rm high}$ due to the weakly coupled bath spins (see Fig.~\ref{FIG:2}), and the fluctuation caused by nuclear spin flipping.
Setting an appropriate threshold photon number (e.g., $n_{\rm th}=2.54\times 10^4$ in the case of Fig.~\ref{FIG:3}f),
one can define the fidelity of the single-shot readout (see Appendix \ref{appendix:D} for more analysis of the photon count distribution and the fidelity).
As shown in the case of Fig.~\ref{FIG:3}, the fidelity of single-shot readout of P1 center orientation is $99.4\%$.

\begin{figure*}
  \includegraphics[width=1\textwidth]{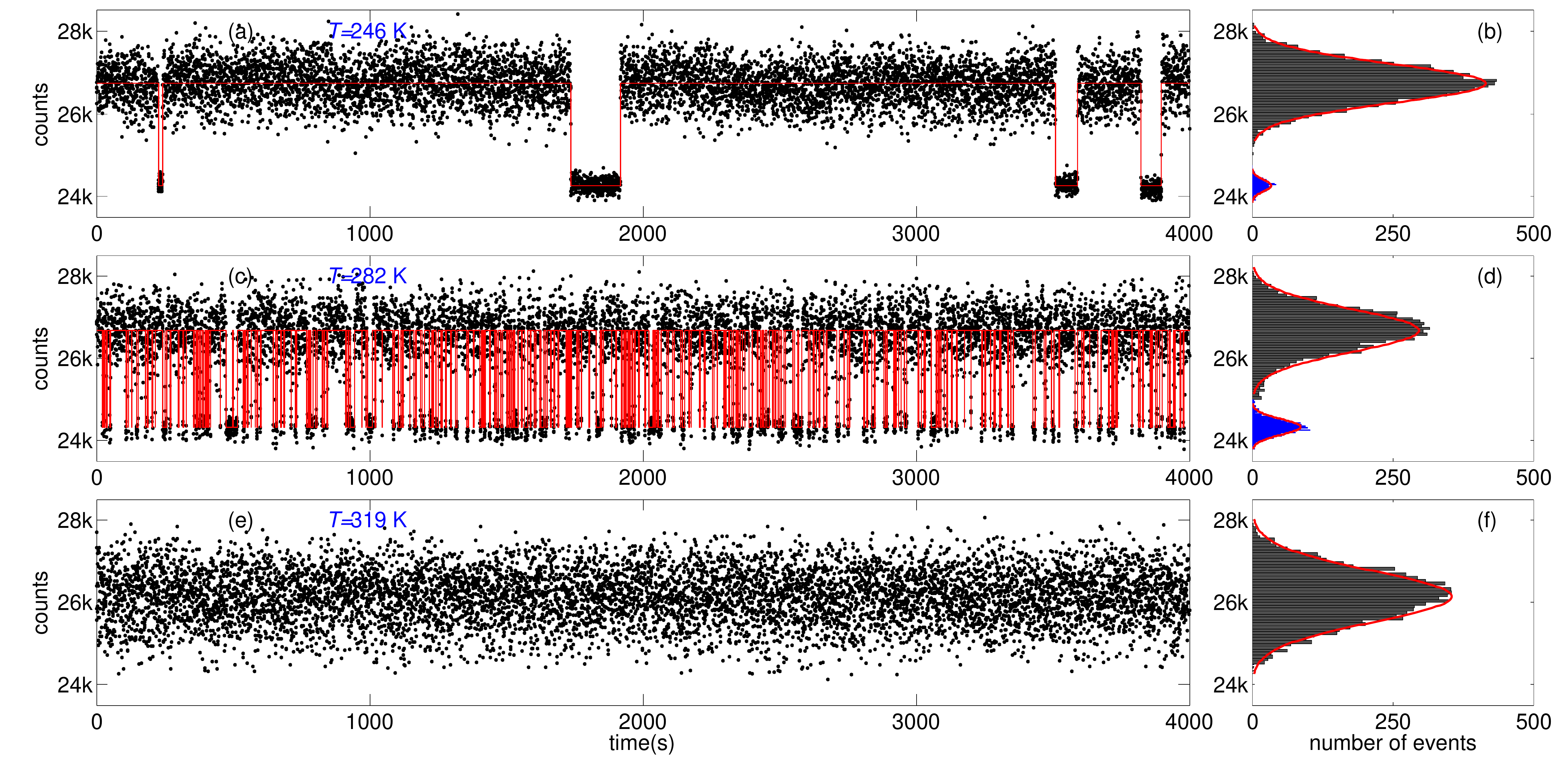}
\caption{(color online) Simulated quantum jumps and single-shot readout of P1 center orientation at different temperatures. 
\textbf{(a)} $\tau_{v}=1000~{\rm s}$ (with $T=246$~K).
\textbf{(b)} Histogram of the photons collected in $0.5~{\rm s}$ of \textbf{a}. The red curves are the Gaussian fitting of the two peaks and
 the fidelity of single-shot readout
of P1 center orientation is 99.8\%.
\textbf{(c)} $\tau_{v}=10~{\rm s}$ (with $T=282$~K).
\textbf{(d)} Histogram of the photons collected in $0.5~{\rm s}$ of \textbf{c}. The red curves are the Gaussian fitting of the two peaks and
 the fidelity of single-shot readout
of P1 center orientation is 94.8\%.
\textbf{(e)} $\tau_{v}=0.25~{\rm s}$ (with $T=319$~K).
\textbf{(f)} Histogram of the photons collected in $0.5~{\rm s}$ of \textbf{e}. Only one Gaussian peak is visible, and the single-shot readout of orientation fails at this temperature. In all simulations nuclear spin relaxation time $T_{1 n}=50~{\rm ms}$ is used..}
\label{FIG:4}       
\end{figure*}

\subsection{Temperature dependence of quantum jumps process} 
The reorientation rate is very sensitive to temperature $T$ [see Eq. (\ref{EQ:1}) ].
With a given nuclear spin relaxation, we investigate the reorientation process 
at different temperatures.
Figure~\ref{FIG:4} shows the simulation results.
The fidelity is decreasing when increasing the temperature, because of the more frequent orientation change during $T_M=0.5~{\rm s}$. If the temperature is high enough (e.g., $T=319$~K), the signal
of JTE will be completely averaged during the time of photon collection. Hence, we expect only one Gaussian peak of the photon counts. Numerical simulation confirms this behavior, as shown in Fig. \ref{FIG:4}f.

\section{Discussion and conclusion}
We have considered the quantum jumps of P1 center with $^{14}$N nuclear spin.
The transition frequencies of four orientations are degenerate when the nuclear spin in the state with $I_z=0$.
This degeneracy prevents us from driving P1 centers with a specific orientation but regardless of their nuclear spin state.
Indeed, the fast nuclear spin relaxation reduces the signal contrast by a factor of $2/3$, in the driving scheme discussed above (Fig.~\ref{EQ:1}c).
This will be different if the bath consists of P1 centers with $^{15}$N nuclear spins.
The degeneracy of transition frequency will be lifted, and readout signal can have a full contrast (see Appendix \ref{appendix:F}).

The above analysis demonstrates the real-time measurement of dynamic JTE of a single P1 center.
In our simulation, we notice that, for randomly generated spin bath, it is possible to have several P1 centers that are strongly coupled to the NV center.
In this case, the NV center spin coherence will be sensitive to the states of these P1 centers.
Using the same DEER sequence demonstrated here, it is possible to observe quantum jumps of more than one P1 centers.

Before conclusion, we point out that our proposal does not strongly depend on the details of detected quantum objects (e.g., their detailed electronic structures).
The single-shot readout measurement will work if (i) the detected object carries either electron spin or nuclear spin, and it couples to the NV center spin; (ii) the reorientation process causes the change of magnetic resonant frequency; and (iii) the readout sequence is fast enough in comparison with the reorientation rate.
It is possible to fulfill these conditions in various systems such as molecular nano-magnets at low temperature.
Detailed analysis of other physical systems is beyond the scope of this paper.
However, we believe that, using shallow NV centers close to diamond surface, people can observe of the dynamic JTE of external single molecules in the near future.

In this work, we propose to measure the dynamic JTE of single P1 defect centers in diamond.
Thanks to the hyperfine interaction with the nitrogen nuclear spins, the defect center orientation is correlated with the magnetic resonant frequency.
Thus, the orientation can be readout by applying a DEER sequence.
Our work extends the ability of NV centers as an outstanding quantum sensor in atomic scale,
particularly, when the proposed method is generalized to detect the vibrational dynamics of single molecules outside diamond.

 \appendix
 \section{Statistics of Random Spin Bath Configurations}
 \label{appendix:A}
\begin{figure}[htbp]
  \includegraphics[width=0.4\textwidth]{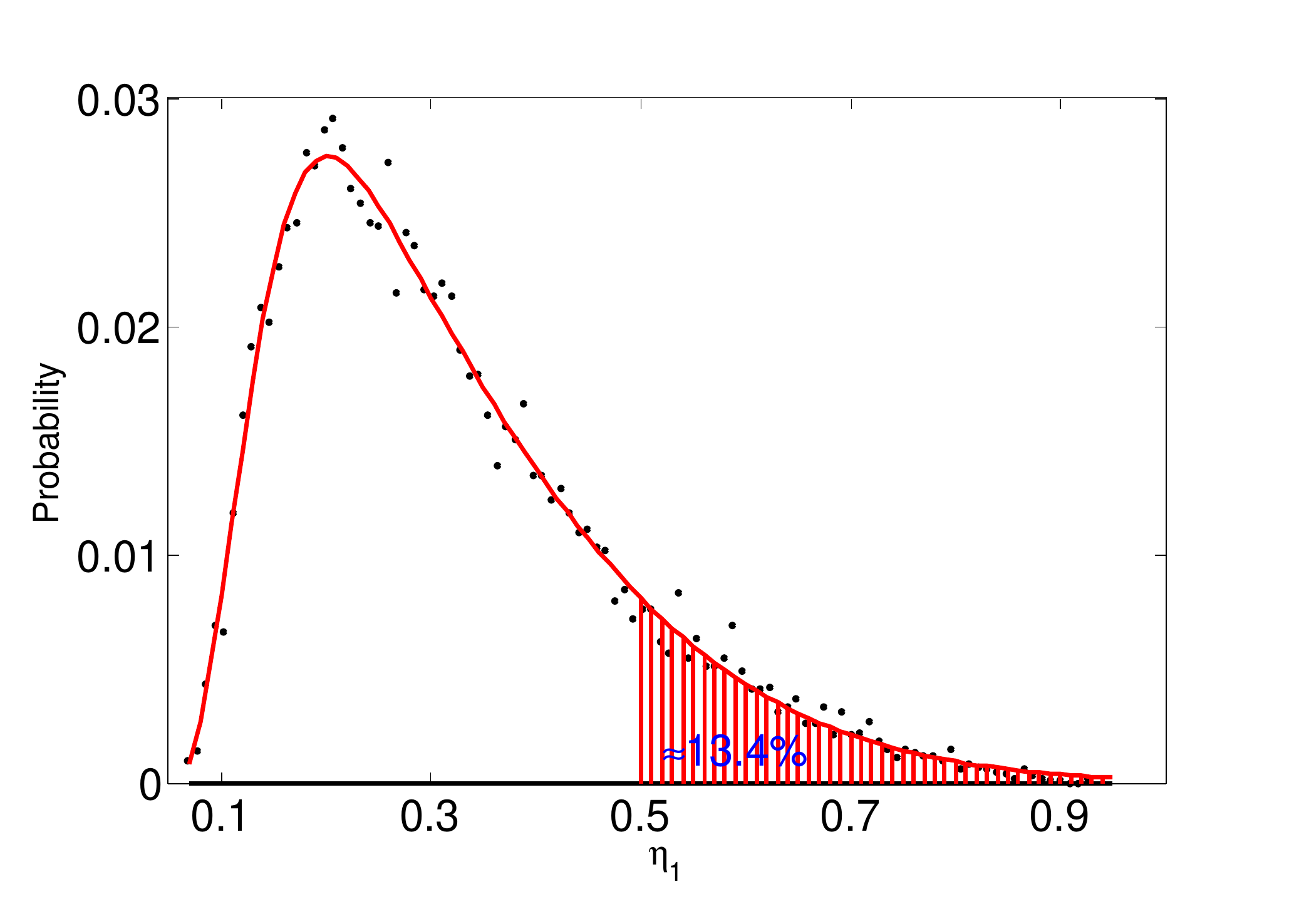}
\caption{(color online) Statistics of the closest P1 center's contribution for $10^4$ randomly generated P1
center spin bath configurations with P1 center concentration $c=200$ ppm. The shadowed area shows that about $13.4\%$ randomly generated bath configurations have a strongly coupled P1 center with parameter $\eta_1>0.5$. }
\label{suppFIG:1}
\end{figure}

In order to observe the JTE, we need the nearest P1 center spin has much stronger interaction with the NV center in comparison with other P1 center spins in the given spin bath configuration.
Since the dipolar coupling strength between P1 center and NV center inverse-cubically
depends on their distance, it is not difficult to find a spin bath configuration
in which the nearest P1 center has a significant contribution to the NV center decoherence, which is characterized by the parameter $\eta_1$ [see Eq.~(\ref{ratio})].
For a typical P1 center concentration $c=200~{\rm ppm}$, we randomly generate  $10^4$ P1
center spin bath configurations and calculate the parameter $\eta_1$ of each configurations.
 Figure S\ref{suppFIG:1} shows that
the probability of the ratio $\eta_{1}>0.5$ is about $13.4\%$.

\section{Calculations of NV Center Spin Coherence under DEER Sequence}
\label{appendix:B}
This section presents the calculation of the NV center electron spin coherence under the DEER sequence.
We will focus on the situation discussed in the main text, where about half of the P1 center bath spins are resonantly flipped by the microwave $\pi$-pulse.
Since the P1 center orientation $v$ and its nuclear spin state $I_z$ are hardly changed in the time scale of NV center decoherence (i.e., $\sim {\rm \mu s}$), the states $\{v, I_z\}$ of each P1 centers are assumed to be static when calculating the coherence.
The effect caused by the jumps of states $\{v, I_z\}$ in a longer time scale is analyzed in the main text.

As discussed in the main text, with given states $\{v, I_z\}$ of each P1 centers and given excitation frequencies of microwave pulse, the bath spins are classified into the resonant group and the off-resonant group (see Fig.~1 of the main text).
Since the microwave pulse on the P1 centers has different effect on the resonant and off-resonant groups, the decoherence due to these two groups are treated differently.

For the resonant group, the physical effect of the $\pi$ pulses on the NV center and P1 centers cancels each other.
In this case, the DEER sequence is indeed equivalent to the free-induction decay (FID, i.e., with only the two $\pi/2$ pulses on the NV center) case.
During the period of several ${\rm \mu s}$, the interaction between bath P1 centers (usually $< 10^2~{\rm kHz}$) in the resonant group  can be neglected, and the decoherence $L_{\rm res}$ caused by the P1 centers in the resonant group is
\begin{equation}
L_{\rm res}(t=2\tau)={\rm Tr}\left[ e^{-i H_{0}^{\rm res} t} e^{i H_{+}^{\rm res} t}\right]=\prod_{k\in {G_{\rm res}}}\cos\left(\frac{b_k t}{2}\right),
\end{equation}
where
\begin{eqnarray}
H_{0}^{\rm res}&=&-\gamma B \sum_{k\in G_{\rm res}} S_k^z,\\
H_{+}^{\rm res}&=&-\gamma B \sum_{k\in G_{\rm res}} S_k^z+ \sum_{k\in G_{\rm res}}   b_{k} S_k^z,
\end{eqnarray}
are the conditional Hamiltonians of P1 center bath spins for the NV center electron spin in two eigen-states of $S_{0}^z$, $\vert m_S=0\rangle$ and $\vert m_S=+1\rangle$, respectively.

For the off-resonant group, the pulses on the P1 centers do not take effect.
Thus, the $\pi$ pulse on the NV center refocuses the static fluctuations due to the P1 center bath spins in the off-resonant group.
In fact, the spin echo of NV center in P1 center electron spin bath has been well-studied.
The noise due to the bath spins in the off-resonant group can be modeled by an Uhlenbeck stochastic process \cite{Lange2010} with correlation function $C(t)=b^2\exp(-\vert t\vert/\tau_{\rm c})$, where $b$ is the characteristic noise strength, and $\tau_{\rm c}$ is the noise correlation time.
With this model, the NV center spin echo signal decays as \cite{Lange2010}
\begin{equation}
L_{\rm off}(t=2\tau)= e^{- R b^2 t^3/12}\equiv e^{-(t/T_2)^3}.
\end{equation}
The typical value of coherence time $T_2=3.5~{\rm \mu s}$ was observed in similar diamond sample (with P1 center concentration about $c \sim 200~{\rm ppm}$) considered in our work.
Notice that, the effective bath spin concentration of the off-resonant group is twice smaller than the full concentration.
The coherence time due to the off-resonant group should be prolonged by a factor of 2, i.e. $T_2\sim 7~{\rm \mu s}$ in this case, since the coherence time is inversely proportional to the bath spin concentration in the dipolar coupled systems \cite{Witzel2010}.
With the single-shot readout with working point $t=1.8~{\rm \mu s}$ used in the main text, the off-resonant group has negligible effect on the NV center spin coherence, i.e., $L_{\rm off}(t=1.8~{\rm \mu s})\approx 1$.

\begin{table*}[htdp]
\tiny
\caption{The $Q$-matrix of the Markov chain simulation of P1 center (with $^{14}\rm N$ nuclear spin) quantum jump. The parameters $\nu$ and $\gamma_{1n}\equiv 1/T_{1n}$ are the relaxation rates of orientation and nuclear spin states, respectively.}
\begin{center}
 \begin{tabular}{c c c c c c c c c c c c c}
    \hline
    \hline
    $Q=(q_{ij})$ & (a,-1) & (a,0) & (a,+1) & (b,0) & (c,0) & (d,0) & (b,-1) & (c,-1) & (d,-1) & (b,+1) & (c,+1) & (d,+1) \\ \hline
    (a,-1) & $-\gamma_{1n}-3\nu$ & $\gamma_{1n}$ & 0 & $0$ & $0$ & $0$ & $\nu$ & $\nu$ & $\nu$ & $0$ & $0$ & $0$ \\
    (a,0) & $\gamma_{1n}$ & $-2\gamma_{1n}-3\nu$ & $\gamma_{1n}$ & $\nu$ & $\nu$ & $\nu$ & $0$ & $0$ & $0$ & $0$ & $0$ & $0$ \\
    (a,+1) & 0 & $\gamma_{1n}$ & $-\gamma_{1n}-3\nu$ & $0$ & $0$ & $0$ & $0$ & $0$ & $0$ & $\nu$ & $\nu$ & $\nu$ \\
    (b,0) & $0$ & $\nu$ & $0$ & $-2\gamma_{1n}-3\nu$ & $\nu$ & $\nu$ & $\gamma_{1n}$ & $0$ & $0$ & $\gamma_{1n}$ & $0$ & $0$ \\
    (c,0) & $0$ & $\nu$ & $0$ & $\nu$ & $-2\gamma_{1n}-3\nu$ & $\nu$ & $0$ & $\gamma_{1n}$ & $0$ & $0$ & $\gamma_{1n}$ & $0$ \\
    (d,0) & $0$ & $\nu$ & $0$ & $\nu$ & $\nu$ & $-2\gamma_{1n}-3\nu$ & $0$ & $0$ & $\gamma_{1n}$ & $0$ & $0$ & $\gamma_{1n}$ \\
    (b,-1) & $\nu$ & $0$ & $0$ & $\gamma_{1n}$ & $0$ & $0$ & $-\gamma_{1n}-3\nu$ & $\nu$ & $\nu$ & 0 & $0$ & $0$ \\
    (c,-1) & $\nu$ & $0$ & $0$ & $0$ & $\gamma_{1n}$ & $0$ & $\nu$ & $-\gamma_{1n}-3\nu$ & $\nu$ & $0$ & 0 & $0$ \\
    (d,-1) & $\nu$ & $0$ & $0$ & $0$ & $0$ & $\gamma_{1n}$ & $\nu$ & $\nu$ & $-\gamma_{1n}-3\nu$ & $0$ & $0$ & 0 \\
    (b,+1) & $0$ & $0$ & $\nu$ & $\gamma_{1n}$ & $0$ & $0$ & 0 & $0$ & $0$ & $-\gamma_{1n}-3\nu$ & $\nu$ & $\nu$ \\
    (c,+1) & $0$ & $0$ & $\nu$ & $0$ & $\gamma_{1n}$ & $0$ & $0$ & 0 & $0$ & $\nu$ & $-\gamma_{1n}-3\nu$ & $\nu$ \\
    (d,+1) & $0$ & $0$ & $\nu$ & $0$ & $0$ & $\gamma_{1n}$ &$0$ & $0$ & 0 & $\nu$ & $\nu$ & $-\gamma_{1n}-3\nu$ \\ \hline \hline
\end{tabular}
\end{center}
\label{Tab:1}
\end{table*}%

Indeed, the bath spins in the off-resonant group can be further decoupled by applying multi-$\pi$-pulse dynamical decoupling sequence, and the coherence time $T_2$ can be, at least,  extended to $10^2~{\rm \mu s}$  \cite{Lange2010}.
If $\pi$ pulses are also applied on the P1 centers simultaneously (a generalized multi-pulse DEER sequence), the  decoherence due to the bath spins in the resonant group is not changed, while the contribution of the off-resonant group can be completely neglected.

The inter-group spin interaction has little effect on the NV center spin coherence.
Because of the large frequency mismatch between the resonant and the off-resonant groups, the inter-group spin pair flip-flop process is greatly suppressed.
Thus, it is reasonable to assume that the parameters characterizing the noise from the off-resonant group (i.e. $b$ and $\tau_{\rm c}$) are not affected  by the spins in the resonant group.
Accordingly, the NV center decoherence is caused independently by the two groups
\begin{equation}
L_{\rm DEER}(t=2\tau)=L_{\rm off}(t)L_{\rm res}(t)\approx L_{\rm res}(t).
\end{equation}

\section{Simulation of Quantum Jump Process}
\label{appendix:C}

The real-time change of each P1 center state is simulated by a continuous-time Markov stochastic process \cite{Anderson1991}.
For P1 centers with $^{14}\rm N$ nuclear spins, the Markov chain $\{X(t),t\in[0, +\infty)\}$ has a twelve-state space
$\mathcal{E}=\{(v,I_{z})\}$ with $v\in\{a, b, c, d\}$ and $I_{z}\in\{-1, 0, +1\}$.
The $Q$-matrix (or infinitesimal generator) of the process is given in Table~\ref{Tab:1}.

The quantum jump of P1 centers is studied by a hold-and-jump process, which is
particularly useful for computer simulation.
For a given P1 center in the bath, at random times $t=  t_{1}, t_{2}, \dots, t_{n}, \dots$, it changes to a new state,
and the sequence of states constitutes a
discrete-time process $S=\{S_n \vert S_{n}\in \mathcal{E}\}$.
Usually, we call $t_{n}$ the jump times and $\tau_{n}=t_{n}-t_{n-1}$ the holding times (with $t_0\equiv 0$).
For example, $\tau_{1}$ is the time that the P1 center stays  in $S_{0}$ before it jumps to $S_{1}$.
The holding times of the $i$th state are random variables follows exponential distribution with mean value $q_{i}=-q_{ii}=\sum_{j\neq i}q_{ij}$, where $q_{ij}$ is the matrix element of the $Q$ matrix.

The following algorithm is performed to implement the hold-and-jump process of each single P1 center in the bath:

\ \ (1) Set a total evolution time $T_{\rm tot}$; start the process at $t=0$ with a randomly generated initial state $i\in \mathcal{E}$;

\ \ (2) For the current state $i$, generate a holding time $\tau$, which follows an exponential distribution with parameter $q_{i}$;

\ \ (3) Replace the value of $t\leftarrow t+\tau$; set the jump matrix $\Pi=(\pi_{ij})$ with $\pi_{ii}=0$ and $\pi_{ij}=q_{ij}/q_{i}$.

\ \ (4) Randomly choose a new state $j$ with the probability distribution given by the \emph{i}th
row of the jump matrix $\Pi$;

\ \ (5) If $t<T_{\rm tot}$, set $i\leftarrow j$ and return to step (2); otherwise, the simulation is completed.

The simulated jump process of the nearest P1 center to the NV center is shown in Fig. 3(a) of the main text. Other P1 centers located at different positions have similar random jump behavior.

\section{Single-shot Readout Fidelity}
\label{appendix:D}
In this appendix, we analyze the fidelity of the single-shot readout process.
To this end, we first study the photon count distribution.

\textbf{Photon count distribution.}
The photon count distribution, in general, is of a double-Gaussian shape. The broadening of the Gaussian peaks comes from three sources: (i) the photon shot noise; (ii) the fluctuation of NV center coherence caused by the weakly coupled P1 centers; and (iii) the $^{14}$N nuclear spin flipping of the detected (nearest) P1 center. 
The influence of these mechanisms on the photon count distribution is discussed as follows.

\textbf{Photon shot noise.} The single-shot readout process is essentially a repetitive measurement on the NV centers. 
With $M$ times independent measurements, one can collect $n$ photons, which is a random variable and follows a normal distribution in the large $M$ limit (i.e.,  $M\gg 1$)
\begin{equation}
Q(n; M, L)=\frac{1}{\sqrt{2\pi  \delta_{ M}^2(L)}} \exp\left[\frac{(n-\bar{n}_{L, M})^2}{2\delta_{M}^2(L)}\right],
\end{equation}
where $\bar{n}_{L, M}$  and $\delta_{M}^2(L)$ is the expectation value and the variance of the photon number for a {\it fixed} coherence $L$ of the NV center.

If the NV center is prepared in $\vert m_S=0\rangle$ state, $M$ readout measurements result in $\bar{n}_{\vert 0\rangle} = M \xi $ photons on average, where a typical value $\xi=0.3$ is used for the mean photon number of each measurement (determined by the count rate, photon collection efficiency, and the duration of the readout laser pulse).
An NV center in $\vert m_S=1\rangle$ state emits less photons than in the $\vert m_S=0\rangle$ state. 
With a contrast factor $C=0.7$, the mean photon number for the $\vert m_S=1\rangle$ state is  $\bar{n}_{\vert 1\rangle}=C\cdot \bar{n}_{\vert 0\rangle}$.
An arbitrary coherence value $L$ is mapped to a mean photon number as
\begin{equation}
\bar{n}_{L, M}
=\frac{1}{2}M\xi\left[(1-C)L +(1+C)\right]
\end{equation}
and the variance is
\begin{equation}
\delta_{M}^2(L) = M \cdot \xi_L (1-\xi_L),
\end{equation}
where $\xi_L=\left[(1-C)L+(1+C)\right]\xi/2$ is the mean photon number per measurement with the given coherence $L$. 

\textbf{Fluctuation due to weakly coupled P1 centers.}
In fact, the coherence $L$ is changing due to the weakly coupled P1 centers. 
As shown in Fig.~\ref{FIG:2}b of the main text, the NV center spin coherence $L_{\rm DEER}$ under DEER sequence follows two normal distribution $P_{\rm low}(L)$ and $P_{\rm high}(L)$ centered at $L_{\rm low}$ and $L_{\rm high}$ and with variances  $\sigma^2_{\rm low}$ and $\sigma^2_{\rm high}$, respectively, i.e.
\begin{eqnarray}
P_{\rm low}(L)&=& \frac{1}{\sqrt{2\pi\sigma^2_{\rm low}}} \exp\left[-\frac{(L-L_{\rm low})^2}{2\sigma_{\rm low}^2}\right], \\
P_{\rm high}(L)&=& \frac{1}{\sqrt{2\pi\sigma^2_{\rm high}}} \exp\left[-\frac{(L-L_{\rm high})^2}{2\sigma_{\rm high}^2}\right].
\end{eqnarray}
As explained in the main text, with a strongly coupled P1 center close to the NV center, and with a properly chosen working point, the two Gaussian peaks are well-separated (i.e. $L_{\rm high}-L_{\rm low} \gg \sigma_{\rm low}+\sigma_{\rm high}$), and their overlap can be neglected.
The fluctuation of the NV center coherence $L$ causes the broadening of the photon count distribution as
\begin{equation}
D_{\rm low/high}\left(n; m\right)=\int Q(n; m, L) \cdot  P_{\rm low/high}(L)dL, 
\end{equation}
where the integral region can be safely extended to $(-\infty, +\infty)$ as long as the coherence distribution $P_{\rm low/high}$ is narrow enough (i.e., $\sigma_{\rm low/high}\ll 1$).

When the P1 center is in the $v=a$ orientation (parallel to the magnetic field), with $M$ measurements (e.g., $M=10^5$), the photon count distribution follows a normal distribution 
\begin{equation}
\label{EQ:Dpara}
D_{\parallel}(n; M)=D_{\rm low}(n; M) \approx \mathcal{N}(\bar{n}_{\parallel}, \Sigma_{\parallel}^2)
\end{equation}
with mean value $\bar{n}_{\parallel}$ and variance $\Sigma_{\parallel}^2$
\begin{eqnarray}
\bar{n}_{\parallel}&=&\frac{1}{2}M\xi\left[(1-C)L_{\rm low} +(1+C)\right],\\
\Sigma_{\parallel}^2&=&\frac{M^2\xi^2(1-C)^2}{4} \sigma_{\rm low}^2+\delta_{M}^2(L_{\rm low}).
\end{eqnarray}
In Eq.~(\ref{EQ:Dpara}), we have neglected the coherence $L$ dependence of the variance $\delta_m^2(L)$ [i.e., $\delta_M^2(L)\approx\delta_M^2(L_{\rm low})$]. This is a good approximation as long as the fluctuation of coherence $L$ is small (i.e., $\sigma_{\rm low/high}\ll 1$), which is the case in this work.

For the P1 center in $v=b, c $ or $d$ orientations (the non-parallel orientations), 
the $^{14}$N nuclear spin flipping will mix the two distributions $D_{\rm low}$ and $D_{\rm high}$.
The coherence $L$ follows the distribution  $P_{\rm low}(L)$ when $I_{z}=0$, 
while it follows $P_{\rm high}(L)$ when $I_{z}=\pm 1$.
For a given measurement number  $M$, $xM$ measurements are performed with $I_z=0$, and $(1-x)M$ measurements are performed with $I_z=\pm 1$, 
where $0<x<1$ is the probability of the nuclear spin in the $I_z=0$ state (see Fig.~\ref{FIG:3}d). 
With a {\it given} value of $x$, the photon number distribution is the convolution of the two distributions $D_{\rm low}$ and $D_{\rm high}$
\begin{equation}
\label{EQ:convolution}
D_{\nparallel}(n; M, x)=\int  D_{\rm low}\left(k;  xM\right)\cdot D_{\rm high}\left(n-k; (1-x)M\right) dk
\end{equation}

Essentially, the probability $x$ itself is a random variable. 
For the moment, we consider the {\it fast relaxation limit} of the $^{14}$N nuclear spin. 
When nuclear spin relaxation time  $T_{1n}$ is much smaller than the measurement time $M T_{\rm DEER}$ (i.e. $T_{1n}\ll MT_{\rm DEER}$),
the nuclear spin flips many times during the time of $M$ measurements.
In this case, three nuclear spin states $I_z=0, \pm 1$ are equally populated, and the fluctuation of $x$ is negligible. 
With the random variable $x$ replaced by its expectation value $\bar{x}=1/3$, 
and with the similar approximation applied in Eq.~(\ref{EQ:Dpara}), 
the distribution $D_{\nparallel}(n; M)$ of the non-parallel case in the fast nuclear spin relaxation limit is  also a normal distribution
\begin{equation}
D_{\nparallel}(n; M)\approx D_{\nparallel}(n; M, \bar{x})\approx \mathcal{N}(\bar{n}_{\nparallel}, \Sigma_{\nparallel}^2),
\end{equation} 
where the mean value $\bar{n}_{\nparallel}$ and the variance $\Sigma_{\nparallel}^2$ are calculated as
\begin{eqnarray}
\bar{n}_{\nparallel}&=&M\xi\left[\frac{1-C}{2}\left(\frac{L_{\rm low}}{3}+\frac{2L_{\rm high}}{3}\right) +\frac{1+C}{2}\right],\\
\Sigma_{\nparallel}^2&=&\frac{M^2\xi^2(1-C)^2}{4}\frac{(\sigma_{\rm low}^2+4\sigma_{\rm high}^2)}{9}\notag\\
&+&\delta_{\frac{M}{3}}^2(L_{\rm low})+\delta_{\frac{2M}{3}}^2(L_{\rm high}).
\end{eqnarray}

\begin{figure}[htp]
  \includegraphics[width=0.35\textwidth]{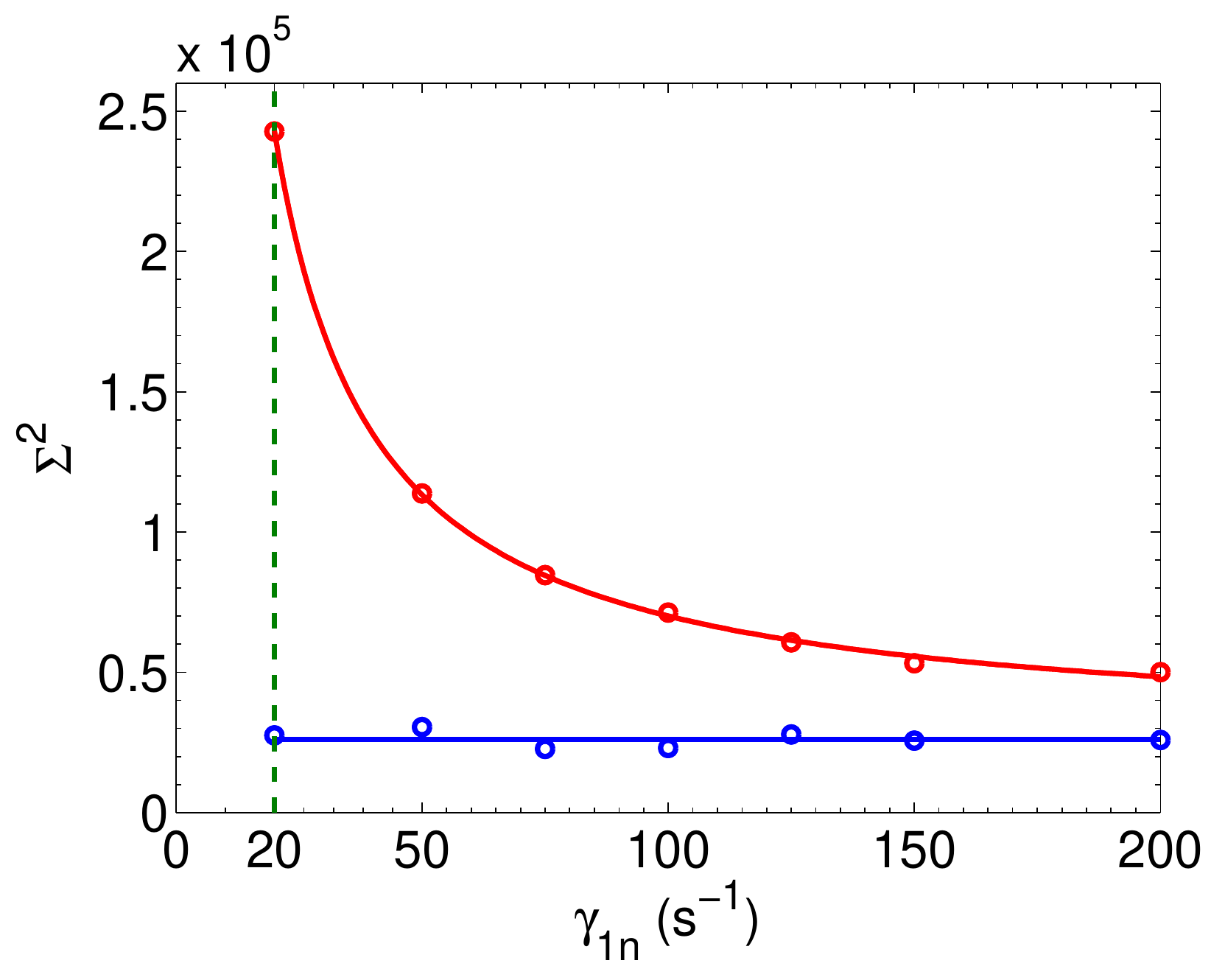}
\caption{(color online)  Simulated photon count variance $\Sigma^2_{\parallel}$ (blue symbols) and  $\Sigma^2_{\nparallel}$ (red symbols).
The curves are fittings of the simulated data. The behavior of the simulated data is consistent with the analysis in the text. 
The variance $\Sigma^2_{\nparallel}$ decays as $\gamma_{1n}^{-1}$, while the variance $\Sigma^2_{\parallel}$ keeps constant.
The simulation are performed with parameters $M=10^5$ and $T_{\rm DEER}=5~\rm{\mu s}$. The vertical dashed line corresponds to the case shown in Fig.~3 of the main text with $\gamma_{1n}=20~\rm{s}^{-1}$.}
\label{suppFIG:9}
\end{figure}

\textbf{Broadening due to nuclear spin flipping.} 
Now, we analyze the effect of fluctuation of $x$ at finite nuclear spin flipping rate  $\gamma_{1n}=1/T_{1n}$. 
In this case, the photon count distribution of the non-parallel case must be averaged over the distribution $W(x)$ of $x$, i.e., 
\begin{equation}
D_{\nparallel}(n; M)=\int_0^1 D_{\nparallel}(n; M, x) \cdot W(x)dx.
\end{equation}
The mean phonon number $\bar{n}_{\nparallel}$ is insensitive to the fluctuation of $x$, 
while the width $\Sigma_{\nparallel}$ is broadened when taking into account the finite variance $s_x^2$ of the distribution $W(x)$.
Our numerical result shows that the variance $s_x^2$ decreases as $s_x^2\propto \gamma_{1n}^{-1}$, which is understandable in the spirit of the central-limit theorem. 
Figure~S\ref{suppFIG:9} demonstrates the simulated photon count distribution variance $\Sigma^2_{\parallel}$ and $\Sigma^2_{\nparallel}$ as functions of nuclear spin relaxation rate $\gamma_{1n}$ with a fixed $M=10^5$.
The variance $\Sigma^2_{\nparallel}$ follows similar behavior as the variance of $s_x^2$ when increasing the relaxation rate $\gamma_{1n}$, while the  variance $\Sigma^2_{\parallel}$ keeps constant.

\begin{figure}[tbp]
  \includegraphics[width=0.4\textwidth]{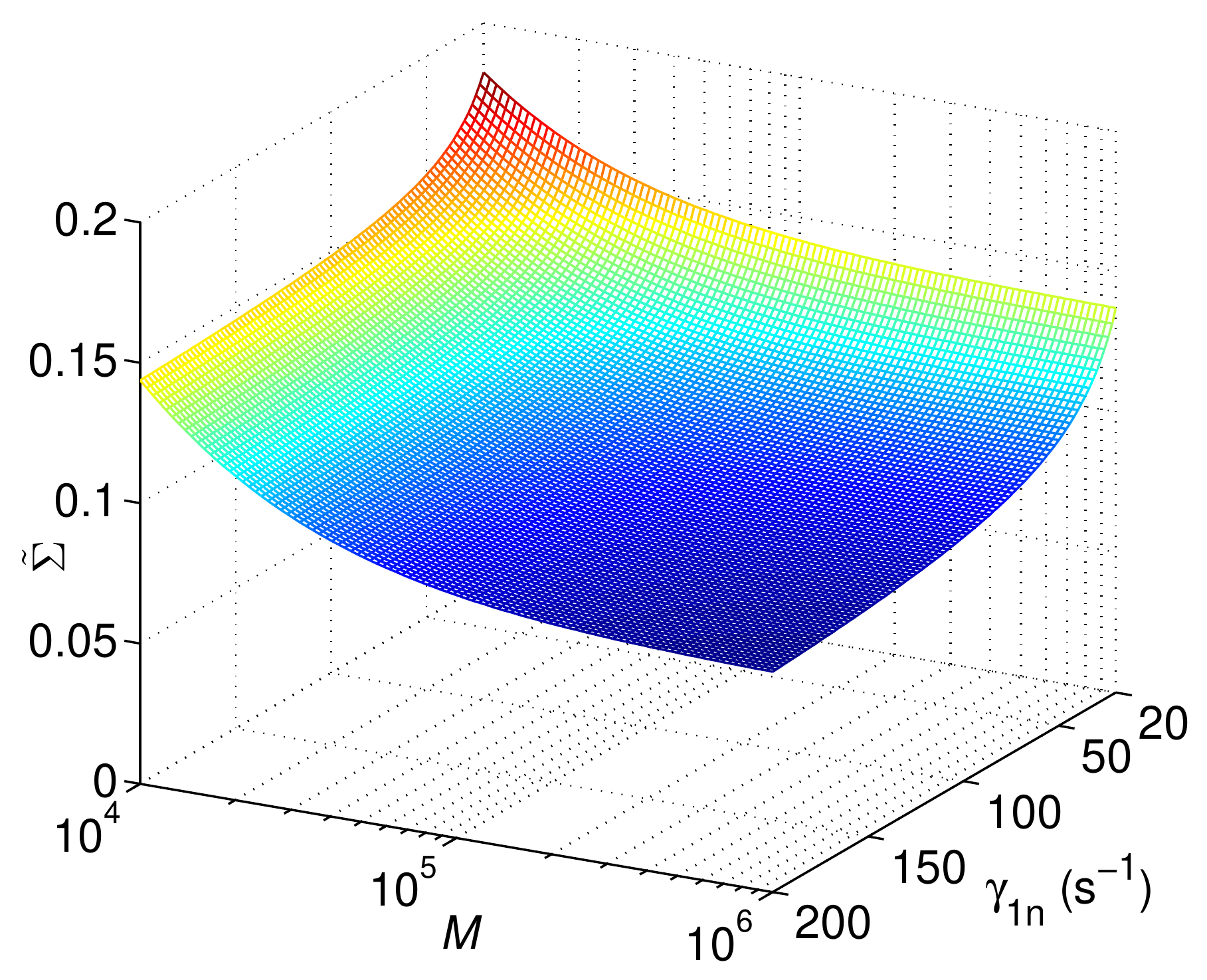}
\caption{(color online)  The normalized photon count distribution widths $\tilde{\Sigma}=\Sigma_{\nparallel} / (\bar{n}_{\nparallel} -\bar{n}_{\parallel})$ as a function of the nuclear relaxation rate $\gamma_{1n}$ and the measurement number $M$. }
\label{suppFIG:8}
\end{figure}
Figure~S\ref{suppFIG:8} shows the width  $\Sigma_{\nparallel}$ (normalized by the peak separation of the two distributions $D_{\rm low}$ and $D_{\rm high}$) as a function of the nuclear relaxation rate $\gamma_{1n}$ and the measurement number $M$.
With a given nuclear spin relaxation $\gamma_{1n}$, increasing measurement number $M$ reduces the relative width,
which is the expected result of decreasing the photon shot-noise and the fluctuation of $x$.

\textbf{Fidelity of Single-Shot Readout.}
In this subsection, we analyze the fidelity of the single-shot readout. 
One obvious reason for the readout error is the overlap of the two Gaussian distributions $D_{\parallel}$ and $D_{\nparallel}$.
Increasing measurement number $M$ will decrease the overlap (see Fig.~S\ref{suppFIG:8}). 
However, another mechanism, namely, the state change of the nearest P1 center during the single-shot readout process (i.e., during the period $T_{M}=MT_{\rm DEER}$), will cause the readout error when increasing the measurement time (i.e., increasing $M$).

To quantify the readout fidelity, we define a threshold $n_{\rm th}$ value of photon number collected with $M$ times repetition measurements as the weighted average of $\bar{n}_{\parallel}$ and $\bar{n}_{\nparallel}$
\begin{equation}
n_{\rm th} \equiv \frac{\Sigma_{\nparallel} \bar{n}_{\parallel}+\Sigma_{\parallel} \bar{n}_{\nparallel} }{\Sigma_{\parallel}+\Sigma_{\nparallel}}.
\end{equation}
The readout fidelity because of the distribution overlap is quantified as
\begin{equation}
F_1=\frac{1}{4}{\rm erf}\left(\frac{n_{\rm th} - \bar{n}_{\parallel}}{\sqrt{2}\Sigma_{\parallel}}\right)+\frac{3}{4}{\rm erf}\left(\frac{\bar{n}_{\nparallel}-n_{\rm th}}{\sqrt{2}\Sigma_{\nparallel}}\right)
\end{equation}
where ${\rm erf}(x)\equiv \frac{2}{\sqrt{\pi}}\int_0^{x} \exp(-t^2)dt$ is the error function.
The pre-factors (i.e., $1/4$ and $3/4$) account for the equilibrium populations of the parallel and the non-parallel orientations.

\begin{figure}[tbp]
  \includegraphics[width=0.4\textwidth]{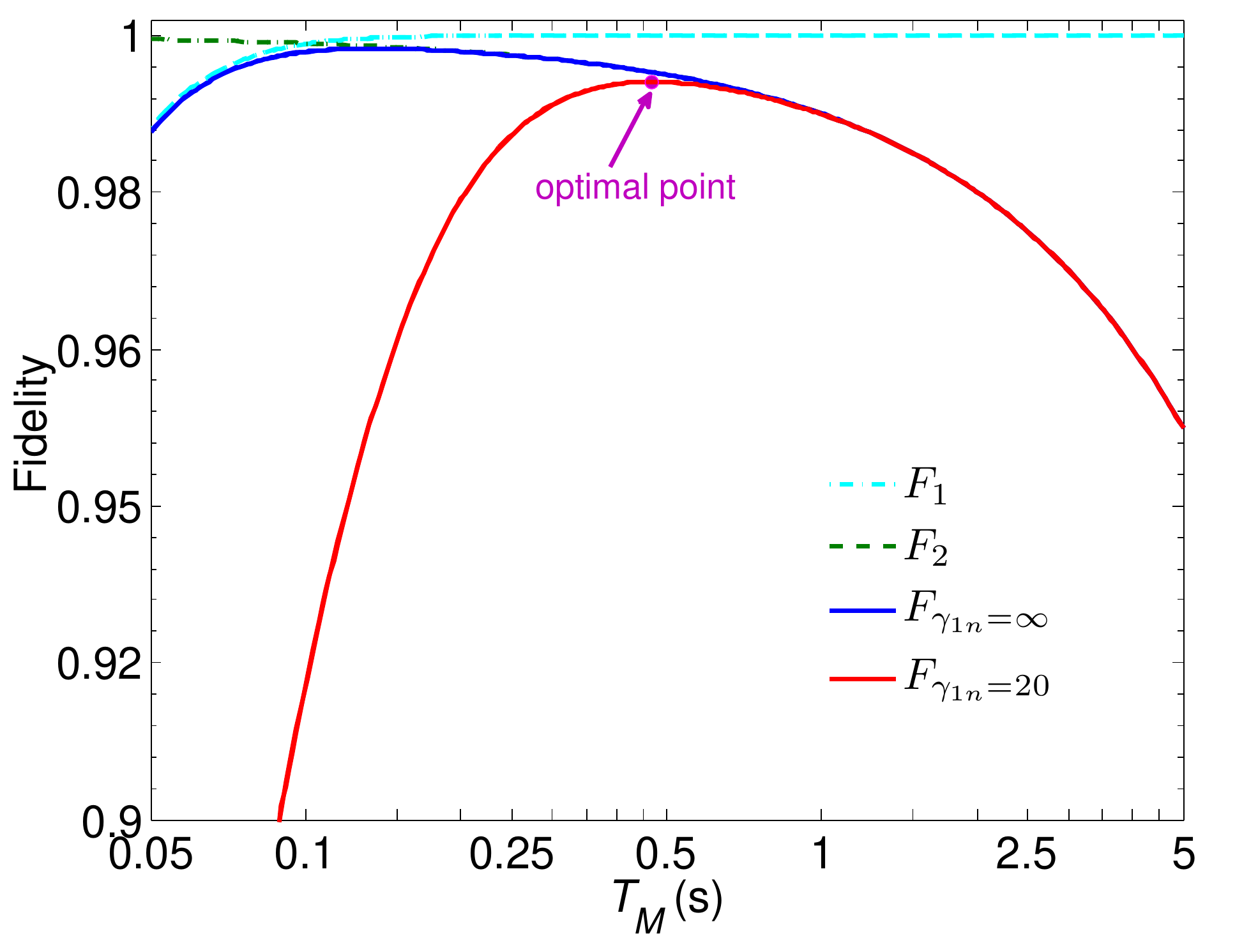}
\caption{(color online)  Single-shot readout fidelity as a function of the measurement period $T_{M}=MT_{\rm DEER}$. 
The cyan, green and blue curves are, in turn, the fidelity components $F_1$, $F_2$ and the total fidelity $F$ in the large $\gamma_{1n}$ limit.
The red curve is the fidelity corresponding to the parameters in the main text (i.e., $\tau_{\nu}=100$ s, $\gamma_{1n}=20~\rm{s}^{-1}$) and working point $t=1.8$ $\mu$s. The fidelity drops significantly when the measurement time $T_M$ approaches the assumed nuclear spin relaxation time $T_{1n}=50~{\rm ms}$ (e.g., the region of $T_M \lesssim 0.1~{\rm s}$). }
\label{suppFIG:5}
\end{figure}

The state change of the detected nearest P1 center during a single measurement period $T_M$ causes additional readout error, which is characterized by the ratio of measurement time $T_M=MT_{\rm DEER}$ to the orientation relaxation time $\tau_{v}$. In this case, the fidelity due to the state change is characterized by
\begin{equation}
F_2=1-\frac{MT_{\rm DEER}}{\tau_{v}}.
\end{equation}

\begin{figure*}[htbp]
  \includegraphics[width=1\textwidth]{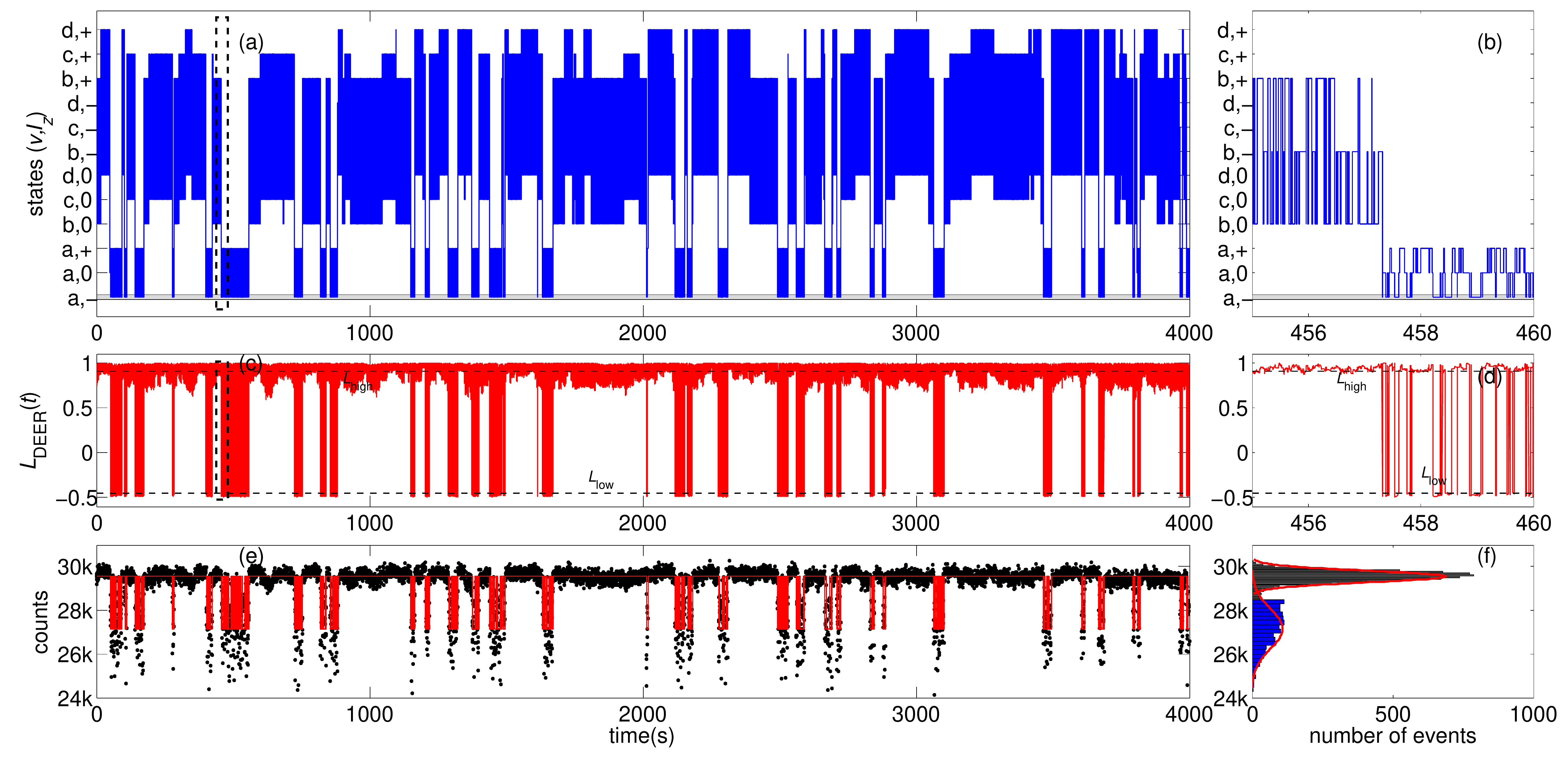}
\caption{(color online)  Single-shot readout of P1 center orientation with single-frequency $\pi$-pulse which only flips the P1 centers in the $(a,-1)$ state. The simulation is performed with reorientation rate $\nu=0.01~{\rm s}^{-1}$ (with $T=262$K) and nuclear spin relaxation time $T_{1 n}=50~{\rm ms}$.
(a) State jumps of the nearest P1 center to the NV center.
(b) Zoom-in of the boxed region in (a).
(c) The calculated NV center spin coherence according to the P1 center states with the working point $t=1.8~{\rm \mu s}$.
(d) Zoom-in of the boxed region in (c).
(e) Simulated quantum jumps of the P1 center orientation. Each data point corresponds to the number of photons collected in $0.5~{\rm s}$ ($10^5$ repetitions of the DEER sequence).
(f) Histogram of the photons collected in $0.5~{\rm s}$. The red curves are the Gaussian fitting of the two peaks.}
\label{suppFIG:3}
\end{figure*}

The total measurement fidelity is the combination of the two components above
\begin{equation}
F=F_1+F_2-1.
\end{equation}

The two components $F_1$ and $F_2$ of the fidelity have different behavior when increasing the measurement time $T_{M}$.
In the short-time case, the readout error is dominated by the overlap between two Gaussian peaks, while
in the large $M$ case,  the frequent change of orientation during the measurement time $T_M$ reduces the fidelity.
The tradeoff of these two errors gives an optimal measurement period, as shown in
Fig.~S\ref{suppFIG:5}.
In the case discussed in the main text, numerical simulation suggests that the optimal measurement scheme is near $M=10^5$ (corresponding to $T_{M}=M T_{\rm DEER}=0.5~{\rm s}$).

\section{Different driving scheme: driving a single resonance $(v=a,I_z=-1)$}
\label{appendix:E}
The reorientation process can be monitored with different driving schemes.
One can drive the bath spins with a single microwave frequency resonant to P1 centers in the state $(v=a,I_z=-1)$.
The spin coherence
switches to $L_{\rm low}$ when the nearest P1 center jumps to the $(a,-1)$ state, otherwise it stays at $L_{\rm high}$.
However, due to the fast relaxation of nuclear spin, the nearest P1 center will quickly switch to different
nuclear spin states and cannot keep in the state $(a,-1)$ for a long time.
Therefore, the photon number collected in 0.5~s has large fluctuation and the signal contrast is also reduced (see Fig. S\ref{suppFIG:3}).
Nevertheless, one can read out the orientation of the nearest P1 center with fidelity $F\sim 80\%$.

\section{P1 Center with $^{15}\rm N$ nuclear spin}

\label{appendix:F}

\begin{figure*}[htbp]
  \includegraphics[width=1\textwidth]{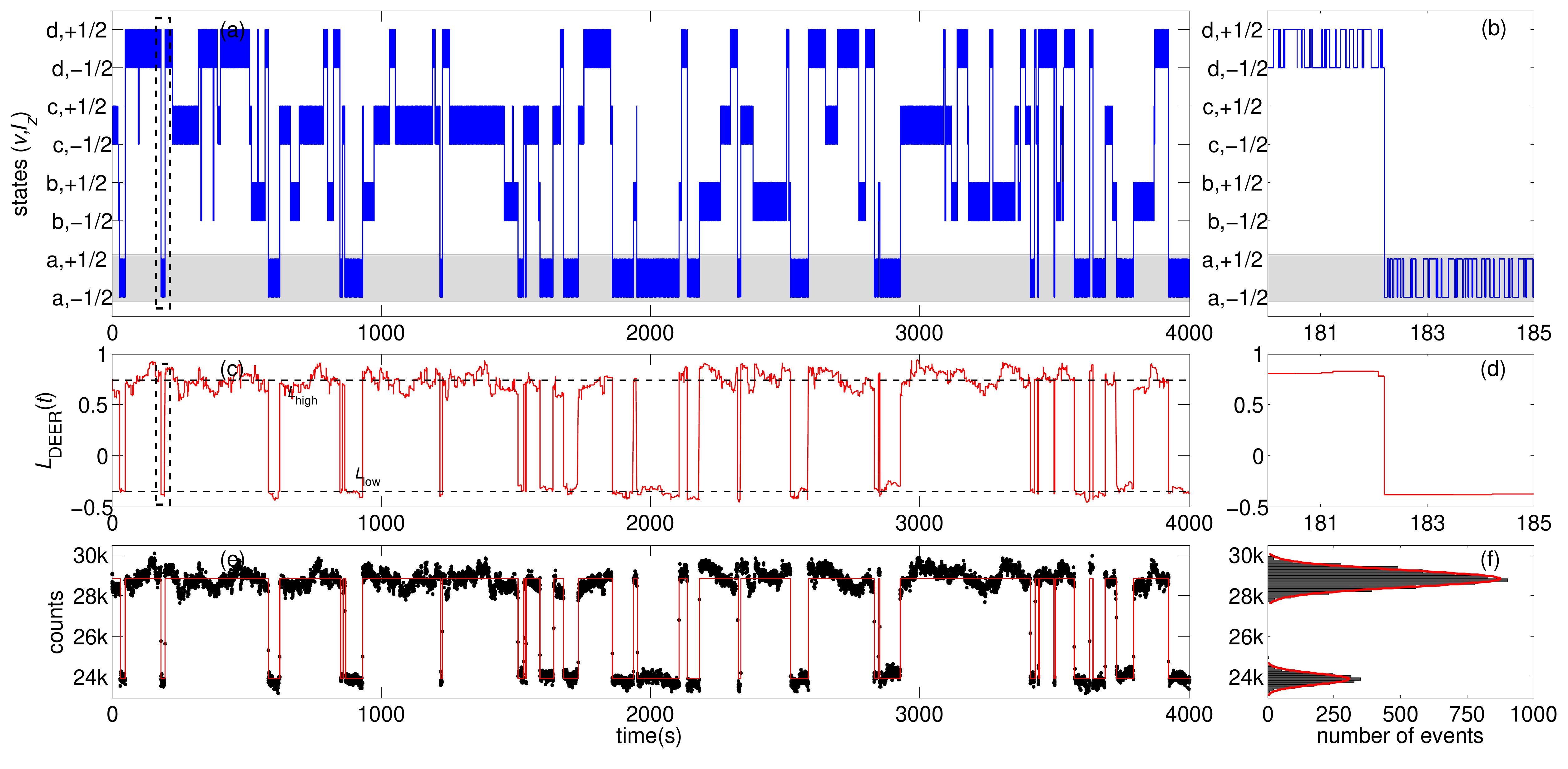}
\caption{(color online)  Result of $^{15}\rm N$ P1 center bath. The simulation is performed with reorientation rate $\nu=0.01~{\rm s}^{-1}$ (with $T=262$K) and nuclear spin relaxation time $T_{1 n}=50~{\rm ms}$.
(a) State jumps of the nearest P1 center to the NV center.
(b) Zoom-in of the boxed region in (a).
(c) The calculated NV center spin coherence according to the P1 center states with the working point $t=1.8~{\rm \mu s}$.
(d) Zoom-in of the boxed region in (c).
(e) Simulated quantum jumps of the P1 center orientation. Each data point corresponds to the number of photons collected in $0.5~{\rm s}$ ($10^5$ repetitions of the DEER sequence).
(f) Histogram of the photons collected in $0.5~{\rm s}$. The red curves are the Gaussian fitting of the two peaks. The fidelity of single-shot readout
of P1 center orientation is 99.5\%.}
\label{suppFIG:4}
\end{figure*}

Due to the degeneracy of four orientations in the state $I_{z}=0$ of $^{14}\rm N$ P1 center,
the signal contrast is reduced by a factor of $2/3$.
If the spin bath consists of P1 centers with $^{15}\rm N$ nuclear spins, four  dips will be observed corresponding to the electron
spins in states $\{(a,\pm1/2)\}$ and $\{(v,\pm1/2)\}$ for $v=b, c$ and $d$.
By driving the two outer peaks (i.e., $\{(a,\pm1/2)\}$), we can define the resonant group $G_{\rm res}$ unambiguously with the P1 center in $a$ orientation and the
off-resonant group $G_{\rm off}$ with the P1 center in $b, c$ and $d$ orientations. In this situation,
the relaxation of nuclear spin states no longer mixes the resonant and off-resonant groups.
Figure S\ref{suppFIG:4} shows the jump process of the nearest $^{15}\rm N$ P1 center, evolution
of the NV center spin coherence, the simulation of single-shot readout of P1 center orientation and the histogram of photons collected in $0.5~{\rm s}$.
Being different with the results in the $^{14}\rm N$ P1 center case, the NV center spin coherence
changes from $L_{\rm high}$ to $L_{\rm low}$ only when the nearest $^{15}\rm N$ P1 center jumps into the $a$ orientation [see the difference between
Fig. 3(d) of the main text and Fig. S\ref{suppFIG:4}(d)]. On the other hand, the signal contrast is improved (to the full contrast of $30\%$) in the  $^{15}\rm N$ P1 center case.

\begin{acknowledgements}
We acknowledge RB Liu's comment on the manuscript and the suggestion for the future work.
We thank XY Pan for the discussions of the DEER measurement.
We thank LM Zhou for the assistance of the numerical simulation.
This work is supported  by NKBRP (973 Program) 2014CB848700 and NSFC No. 11374032, No. 11247006 and No. 11121403.
\end{acknowledgements}


\begin{thebibliography}{22}
\bibitem{Note} The results obtained in this paper are not sensitive to the value of nuclear spin life time $T_{1n}$, as long as the condition $T_{1n}\ll M T_{\rm DEER}$ is satisfied. On the other hand, if the nuclear spin life time is long enough (e.g., in very strong magnetic fields or at very low temperature),  in the opposite limit $T_{1n}\gg M T_{\rm DEER}$, it is possible to monitor quantum jumps of both nuclear spin flipping and reorientation of the strongly coupled P1 center using the same technique. We analyze the nuclear spin flipping effect in detail in the Appendix \ref{appendix:D}.
\end{thebibliography}

%

\end{document}